\documentclass[aps,prb,showpacs,amsmath,amssymb,twocolumn,superscriptaddress]{revtex4-1}
\pdfoutput=1

\usepackage{graphicx}% Include figure files
\usepackage{color}

\renewcommand{\d}{\mathrm{d}}
\newcommand{\ii}{\mathrm{i}}

\newcommand{\mr}[1]{\mathrm{#1}}
\newcommand{\refcite}[1]{Ref.~\onlinecite{#1}}

\begin{document}

\title{Giant current fluctuations in an overheated single electron transistor}

\author{M.~A.~Laakso}
\email[]{matti.laakso@tkk.fi}
\author{T.~T.~Heikkil\"a}
\affiliation{Low Temperature Laboratory, Aalto University, Post Office Box 15100, FI-00076 AALTO, Finland}
\author{Yuli V.~Nazarov}
\affiliation{Kavli Institute of Nanoscience, Delft University of Technology, 2628 CJ Delft, The Netherlands}
\date{\today}

\begin{abstract}
Interplay of cotunneling and single-electron tunneling in a thermally isolated single-electron transistor (SET) leads to peculiar overheating effects. In particular, there is an interesting crossover interval where the competition between cotunneling and single-electron tunneling changes to the dominance of the latter.  In this interval, the current exhibits anomalous sensitivity to the effective electron temperature of the transistor island and its fluctuations. We present a detailed study of the current and temperature fluctuations at this interesting point. The methods implemented allow for a complete characterization of the distribution of the fluctuating quantities, well beyond the Gaussian approximation. We reveal and explore the parameter range where, for sufficiently small transistor islands, the current fluctuations become gigantic. In this regime, the optimal value of the current, its expectation value, and its standard deviation differ from each other by parametrically large factors. This situation is unique for transport in nanostructures and for electron transport in general. The origin of this spectacular effect is the exponential sensitivity of the current to the fluctuating effective temperature.
\end{abstract}


\maketitle

\section{Introduction}\label{sec:intro}

By its statistical physics definition, the temperature of an open electron system in equilibrium is fixed, with a value equal to the temperature of the reservoir it is connected to.\cite{reichl98} The energy of the system can still fluctuate because of the constant exchange of energy with its surroundings. Out of equilibrium, the entire concept of a temperature becomes ill-defined. If the internal relaxation in the system is strong enough, however, the electron energy distribution still follows the equilibrium Fermi function with some effective temperature that is determined by a balance of the energy currents flowing into and out of the system.\cite{giazotto06} In this out-of-equilibrium situation the intrinsic fluctuations of the energy currents translate directly to fluctuations in the effective temperature of the system, provided that the electron--electron relaxation time is significantly shorter than the energy relaxation time to the reservoirs or the phonon bath.\cite{heikkila09} If the electron--phonon interaction in the system is weak, the dominant energy flows are to the reservoirs which are used to drive the system out of equilibrium. In this case the system is said to be fully overheated. As the size of the system becomes ever smaller, full overheating becomes easier to achieve, or, harder to avoid.

One example of a system where overheating is important to take into account is the single-electron transistor (SET).\cite{averin86} When the size of the Coulomb island is decreased ever further in the pursuit for sensitivity, the temperature of the island begins to be affected by the electronic heat flows to the surroundings, e.g., leads.\cite{korotkov94,liu97,sukhorukov01} In an earlier paper\cite{laakso10} we studied the fully overheated single-electron transistor, and found out that the inclusion of inelastic cotunneling,\cite{averin90} along with the often dominant sequential tunneling of single electrons, is important and leads to peculiar overheating effects. The electron transport through the overheated SET is divided into three regimes: cotunneling dominated, competition, and single-electron dominated regimes. We also found that the electric current noise in the crossover interval is mostly contributed by the frequency interval corresponding to typical time scale of temperature equilibration. This is because of the anomalously strong temperature sensitivity of the current in this interval, which makes the temperature fluctuations directly visible in the current (noise).

\begin{figure}
 \centering
 \includegraphics[width=0.8\columnwidth]{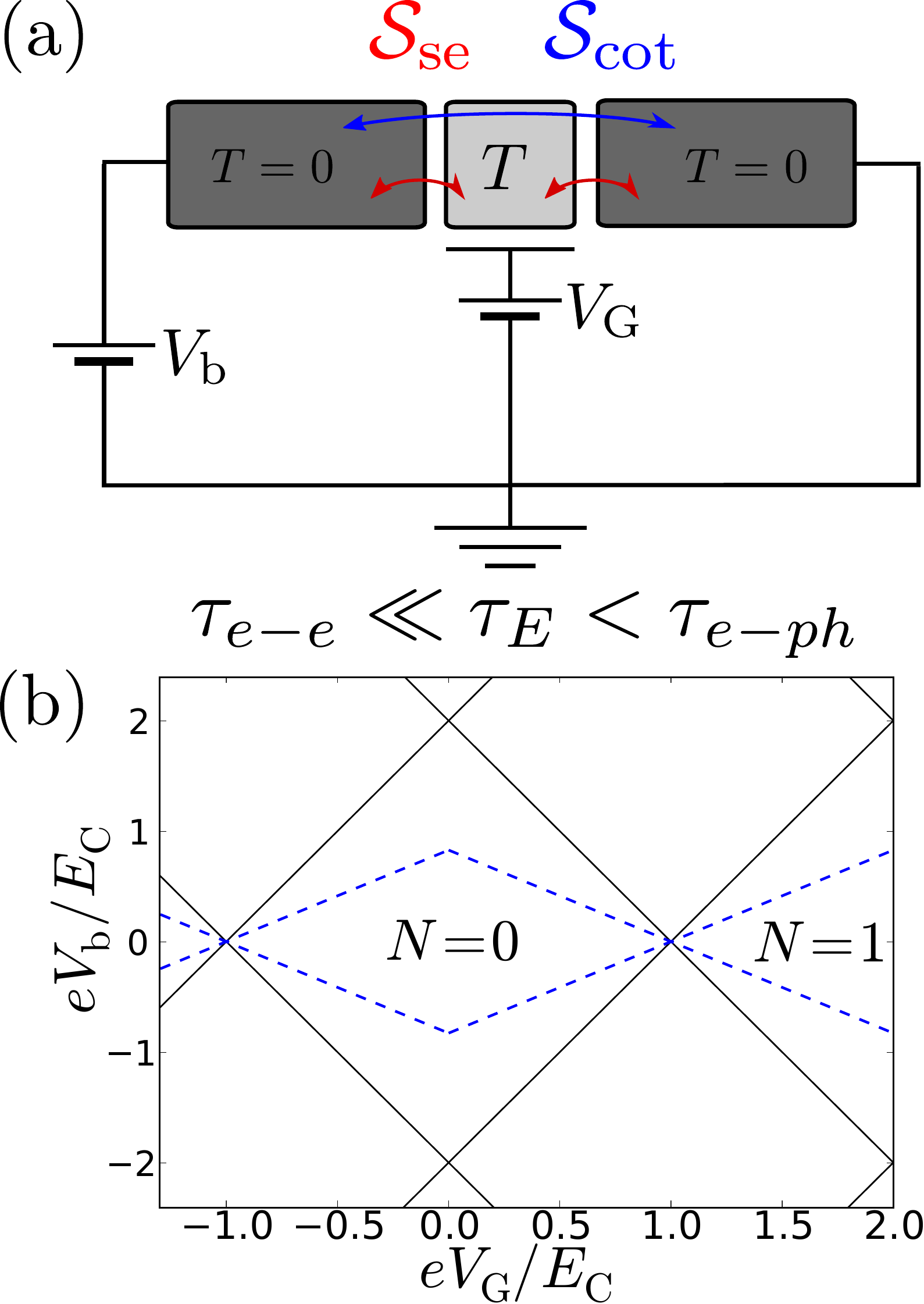}
 \caption{(color online) (a) A schematic diagram of a single-electron transistor, biased by voltage $V_b$. The charge on the island can be tuned with the gate voltage $V_G$. (b) Coulomb diamonds in a symmetric SET. Blue dashed line shows the threshold voltage $V_C$. Changing the gate voltage has the effect of changing $V_C$.}
 \label{fig:schema}
\end{figure}

In this Article we explore and quantify further the temperature fluctuations and the associated current fluctuations in an overheated SET. We study the SET under conditions of well-developed Coulomb blockade, so that the junction conductances $G_{T,L},\;G_{T,R}\ll e^2/\hbar$. We assume that the level spacing on the island is small compared to its temperature and the charging energy, $\delta_I \ll T\ll E_C\equiv e^2/2C$, which is the case in metallic SETs. To have a clear model, we assume a vanishing temperature in the leads.\footnote{In practice, this means the lead temperature should satisfy $T_L \ll T_C$, where $T_C$ is specified in Eq.~\eqref{eq:t0}.} Since a possible asymmetry of the SET does not affect our results qualitatively, we assume $G_{T,L}=G_{T,R}=G_T$ and define a dimensionless conductance $g_T\equiv G_T\hbar/e^2\ll 1$. Our focus is mainly on the fluctuations around the crossover between the competition and single-electron regimes, $V_b\approx V_C=(\sqrt{2}-1)V_\mr{th}$, i.e., well below the zero temperature Coulomb blockade threshold, $eV_\mr{th}=2(E_C-eV_G)$ (see Fig.~\ref{fig:schema}). Note that in the actual experimental setup $eV_G$ must be replaced with a combination of the source, drain, and gate voltages together with the respective mutual capacitances.\cite{ingold92} Here we lump all this into one gate voltage dependence. The electron--electron relaxation time, $\tau_{e-e}$, is assumed short compared to the characteristic time scale for the energy relaxation to the leads, $\tau_E$, which in turn is assumed small compared to the electron--phonon relaxation time, $\tau_{e-ph}$.\footnote{We estimate $1/\tau_{e-e} \simeq R_i G_Q T/\hbar$, $R_i$ being the resistance of the island not including the tunnel barriers. The ratio of times $\tau_E/\tau_{e-e}$ is then estimated as $R_i G_Q g_T (T/\delta_I)$. In practice, it is always large for not extremely pure islands. For the electron--phonon coupling, $\tau_E/\tau_{e-ph}=\hbar\Sigma\mathcal{V}T_C^3/k_B^2g_T$, where $\mathcal{V}$ is the volume of the island and $\Sigma$ the material-specific electron--phonon coupling constant.} These conditions ensure that the island is fully overheated and in quasi-equilibrium. 

\begin{figure}
 \centering
 \includegraphics[width=\columnwidth]{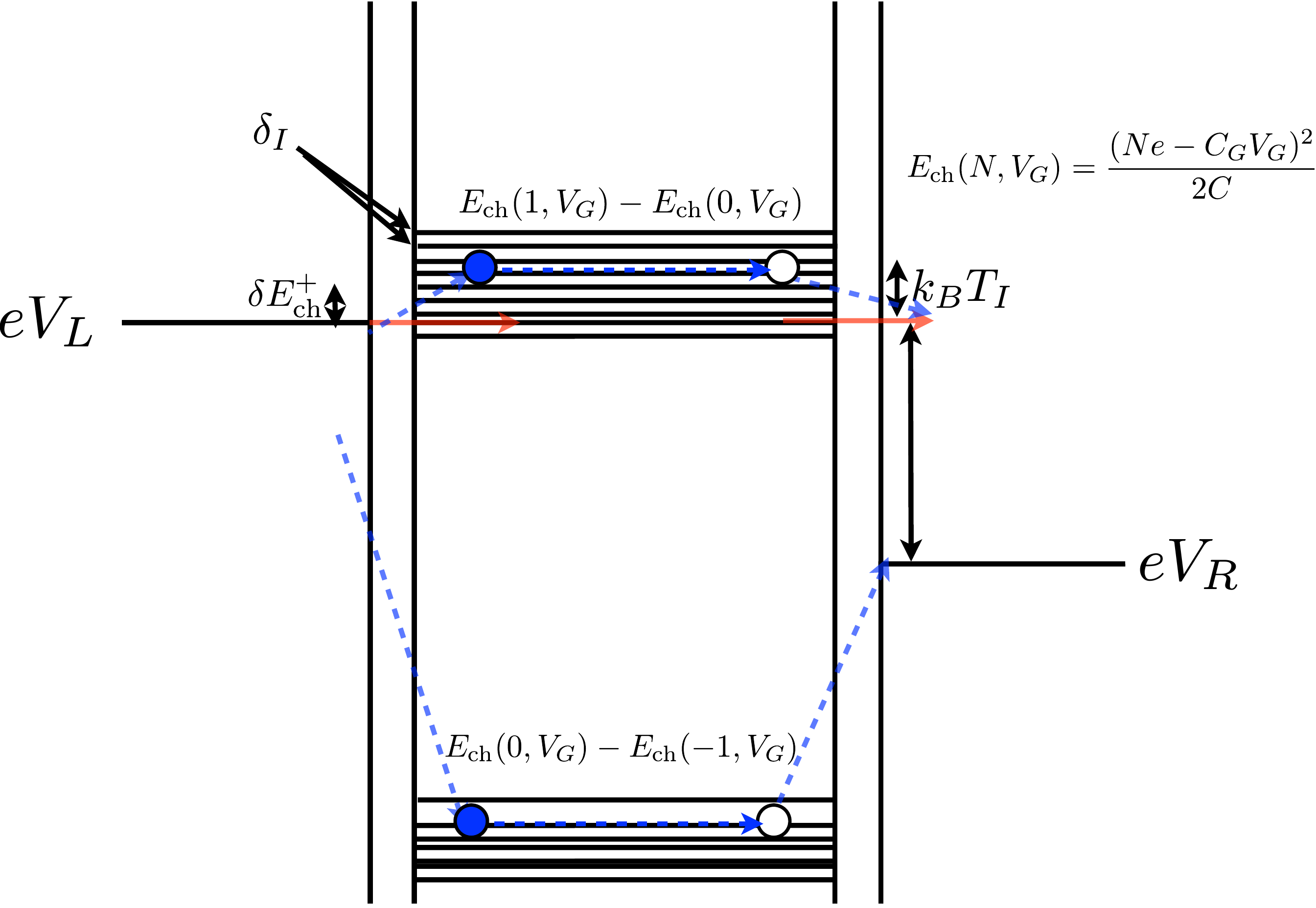}
 \caption{(color online) Schematic energy balance of the single-electron transistor within the Coulomb blockade regime, where the allowed charge state energies are outside the bias window. In this case the temperature of the island is fixed by the balance between sequential tunneling (solid red arrows) and inelastic cotunneling (dashed blue arrows), where the strength of the previous depends on the exponentially small ($\propto \exp(-\delta E^+/k_B T)$) amount of empty (filled) single-electron states below (above) the Coulomb blockade threshold corresponding to the addition (removal) of an electron to (from) the island. This leads to the extreme temperature sensitivity of the current, which is the main topic of this paper.}
 \label{fig:energydiagram}
\end{figure}

Within the Coulomb blockade, the temperature of the island is fixed by a balance between energy flows in and out of the island due to inelastic cotunneling and sequential tunneling (see Fig.~\ref{fig:energydiagram}). Inelastic cotunneling leaves behind an electron--hole excitation on the island, always heating it up. Sequential tunneling can either cool or heat the island, depending on bias voltage. As discussed in Ref.~\onlinecite{laakso10}, there are three qualitatively different regions of bias voltages, revealed by the behavior of the average temperature: At the lowest voltages, sequential tunneling is completely suppressed and the average temperature is fixed by the inelastic cotunneling. Close to the Coulomb blockade threshold (and above), inelastic cotunneling effects can be disregarded and sequential tunneling sets the temperature. The most interesting region is the intermediate regime where the competition between these two processes gives rise to both anomalously sensitive temperature dependence of the current and to strongly non-Gaussian temperature fluctuations. In this regime, we find (see Sec.~\ref{sec:theory}.A and App.~\ref{sec:rescaling}) for the average temperature $\langle T \rangle \approx T_C$ where $T_C$ satisfies 
\begin{equation}
\alpha g_T\left(\frac{V_C}{\sqrt{2}T_C}\right)^{3} \exp\left(\frac{1}{\sqrt{2}}\left(\frac{V_C}{T_C}-1\right)\right)=1,
\label{eq:t0}
\end{equation}
$\alpha$ being a dimensionless coefficient, $\alpha\approx 0.1$. For example, for $g_T=10^{-3}$, $T_C/V_C \approx 0.1$ and for smaller values of $g_T$, $T_C/V_C \approx 1/[\sqrt{2}\ln(1/g_T)]$. As we show below, the small value of $T_C/V_C$ allows us to rigorously describe the anomalously large and strongly non-Gaussian temperature and current fluctuations in an overheated SET.

We find that the probability distributions of the temperature and current fluctuations are strongly non--Gaussian, and that the size of the transistor island has a large influence on these distributions. In particular, we find that for small islands with relatively large single-electron level spacing,
\begin{equation*}
 \delta_I\gtrsim\left(\frac{T_C}{V_C}\right)^5V_C,
\end{equation*}
the mode, expectation value, and standard deviation of the temperature and current differ from each other by parametrically large factors (see Figs.~\ref{fig:avgy} and \ref{fig:avgj}, and Eqs.~\eqref{eq:modej}, \eqref{eq:avgj} and \eqref{eq:varj}). This we show by first casting the Keldysh action technique used in Refs.~\onlinecite{heikkila09} and \onlinecite{laakso10} into the form of a Fokker--Planck equation, allowing us to describe the full normalized probability distribution, instead of the non-normalized large deviation function $\ln P$. For some realistic values, $T_C/V_C\approx0.1$ and $E_C\approx1\:\mr{meV}$, $\delta_I\gtrsim10^{-8}\:\mr{eV}$, corresponding to roughly $10^9$ atoms.\cite{nazarov09}

This Article is structured as follows: Section \ref{sec:theory} introduces the theoretical methods utilized in this paper. In Sec.~\ref{sec:small} we concentrate on the regime of small fluctuations and in Sec.~\ref{sec:large} on fluctuations in systems with relatively large single-electron level spacing. We conclude and discuss our findings in Sec.~\ref{sec:conclusion}. In most of the text, we use units such that $\hbar=k_B=e=1$, but restore them in the important results.

\section{Approach}\label{sec:theory}

We use the extended Keldysh action technique\cite{kamenev09} as the basis of our theoretical calculations. Within this framework it is straightforward to calculate the full statistics of, for example, electric and energy current.\cite{nazarov02,kindermann04} The effective action of the system is augmented with counting fields $\chi_{L(R)}(t)$ and $\xi_{L(R)}(t)$ for charge and energy transfer to the left (right) reservoir, respectively. The probability to obtain a certain realization of fluctuating energy currents is
\begin{align}
 &P[\dot{H}_L(t)]P[\dot{H}_R(t)]=\int\mathcal{D}\xi_L(t)\mathcal{D}\xi_R(t) \nonumber \\ &\times\exp\left\{-\int\d t\left(\ii\xi_L\dot{H}_L+\ii\xi_R\dot{H}_R-\mathcal{S}(\xi_L,\xi_R,E)\right)\right\},
\end{align}
where $\mathcal{S}$ is the effective action describing the transport processes.\cite{bagrets03} Analogous expression gives the probabilities for charge currents with the substitution $\xi\mapsto\chi$. To ensure the conservation of energy, $E$, on the island, i.e., $\dot{H}_L(t)+\dot{H}_R(t)=\dot{E}(t)$, we use a delta functional,\cite{pilgram04}
\begin{align}
 &\delta[\dot{H}_L(t)+\dot{H}_R(t)-\dot{E}(t)]=\int\mathcal{D}\xi(t) \nonumber \\ &\times\exp\left\{\ii\int\d t\xi\left(\dot{H}_L+\dot{H}_R-\dot{E}\right)\right\},
\end{align}
to generate the conditional probability
\begin{align}
 &P[\dot{H}_L(t),\dot{H}_R(t)]=\int\mathcal{D}\xi_L(t)\mathcal{D}\xi_R(t)\mathcal{D}\xi(t)\mathcal{D}E(t) \nonumber \\ &\times\exp\left\{-\int\d t\left(\ii\left(\xi_L-\xi\right)\dot{H}_L+\ii\left(\xi_R-\xi\right)\dot{H}_R\right.\right. \nonumber \\ &\qquad\quad\biggl.\left.+\ii\xi\dot{E}-\mathcal{S}(\xi_L,\xi_R,E)\right)\biggr\}.
\end{align}
In the case of the single-electron transistor adding a delta functional for the conservation of charge is not necessary, since the effective action we use already conserves charge. Integrating over $\dot{H}_L$ and $\dot{H}_R$ yields the energy-conserving Keldysh partition function as
\begin{equation}\label{eq:keldyshZ}
 \mathcal{Z}=\int\mathcal{D}\xi(t)\mathcal{D}E(t)\exp\left\{-\int\d t\left(\xi(t)\dot{E}(t)-\mathcal{S}(\xi,E)\right)\right\},
\end{equation}
where the imaginary unit has been absorbed into $\xi$.

To calculate the probability of the island having energy between $E^\ast\ldots E^\ast+\Delta E$ during time $\tau\ldots\tau+\Delta\tau$ the functional integration over $E(t)$ in Eq.~\eqref{eq:keldyshZ} is performed over paths which satisfy this requirement. In general, the functional integration cannot be performed analytically. The variation of the exponent with respect to $E$ and $\xi$ yields the semiclassical equations of motion,
\begin{equation}\label{eq:semiclassical}
\dot{E}=\partial_\xi \mathcal{S}, \quad \dot{\xi}=-\partial_E
\mathcal{S},
\end{equation}
which, when solved with the desired boundary conditions $E(\tau)=E^\ast$ and $\xi(-\infty)\to0$, give the saddle point trajectories of $E$ and $\xi$. The probability is then obtained as the saddle point value of the functional integral.\cite{heikkila09} The saddle point approximation is valid when the fluctuations around the semiclassical trajectories are small. In our case this translates to the condition $\delta_I\ll(T_C/V_C)^3V_C$.

For another approach, we expand $\mathcal{S}$ to second order in $\xi$. Then the Gaussian integral over $\xi$ in Eq.~\eqref{eq:keldyshZ} can be performed and the remaining functional integral transformed into a Fokker--Planck equation for the time evolution of the probability distribution of $E$:\cite{kamenev09} 
\begin{align}\label{eq:fp}
 \frac{\partial\mathcal{P}(E,t)}{\partial t}&=\hat{H}\mathcal{P}(E,t), \nonumber \\
 &\equiv-\frac{\partial}{\partial E}\left[D_1(E)\mathcal{P}(E,t)-D_2(E)\frac{\partial\mathcal{P}(E,t)}{\partial E}\right].
\end{align}
Here, the operator $\hat{H}$ is obtained from $\mathcal{S}(\xi,E)$ with the substitution $\xi\mapsto-\partial/\partial E$. Before applying this ``quantization rule'', the operator $\hat{H}$ must be normally ordered, i.e., $\xi$ must be to the left of all $E$ (see Appendix \ref{sec:fp}). The Fokker--Planck equation also takes into account the Gaussian fluctuations around the saddle point trajectory, and is therefore more accurate than the saddle point approximation.

The stationary probability distribution satisfies $\hat{H}\mathcal{P}_\mr{st}(E)=0$, and has the form 
\begin{equation}\label{eq:stationary}
 \mathcal{P}_\mr{st}(E)=\mr{const.}\times\exp\left(\int\d E\frac{D_1(E)}{D_2(E)}\right),
\end{equation}
provided that $\int\d E(D_1(E)/D_2(E))\to-\infty$ when $E\to\pm\infty$. The normalization condition $\int\d E\mathcal{P}_\mr{st}(E)=1$ fixes the value of the constant prefactor.

The time dependence of $E$ can also be described in terms of a Langevin equation. The Fokker--Planck equation is equivalent to a Langevin equation\cite{altland10}
\begin{equation}\label{eq:langevin}
 \frac{\partial E(t)}{\partial t}=D_1(E(t))+\frac{\partial D_2(E(t))}{\partial E(t)}+\eta(t),
\end{equation}
where $\eta(t)$ is a random ``force'' with mean $\langle\eta(t)\rangle=0$ and variance $\langle\eta(t)\eta(t')\rangle=2D_2(E(t))\delta(t-t')$.

To obtain a probability distribution for the effective electron temperature, we must apply a model which describes the relation between $E$ and $T$. We first start by pointing out that the energy fluctuations show up as fluctuations of the electron distribution function $f(\epsilon)$, where $\epsilon$ is the energy of a given excitation. In a free-electron model, the total thermal energy on the island is related to this via 
\begin{equation}
 E(t)=\frac{1}{\delta_I}\int\d\epsilon \epsilon (f(\epsilon,t)-f_0(\epsilon)),
 \label{eq:thermalenergy}
\end{equation}
where $\delta_I$ is the single particle level spacing and $f_0(\epsilon)$ is the zero-temperature electron distribution function. In the quasiequilibrium limit, $f(\epsilon)$ tends to a Fermi distribution function, $f(\epsilon,t)=(\exp\{\epsilon/[T(t)]\}+1)^{-1}$, with a fluctuating electron temperature. In this case we can integrate Eq.~\eqref{eq:thermalenergy} and obtain the usual expression for the free-electron heat capacity $C(T)$, 
\begin{equation}\label{eq:energy}
 E=\frac{C(T)T}{2}=\frac{\pi^2T^2}{6\delta_I}.
\end{equation}
Note that in principle the distribution of the fluctuating electron temperature may depend on the charge state $n$. However, under Coulomb blockade the relaxation rate of the excited charge states is larger than the temperature equilibration rate by a large factor, of the order of $E_C/\delta_I$, and this effect can be disregarded in the following.

\subsection{Single-electron transistor}

For the overheated single-electron transistor the SE part of the action in the limit $T\ll V_b,V_C$ reads\cite{laakso10}
\begin{align}
\label{eq:seqaction}
 {\cal S}_{\mr{se}}=&-g_T Te^{-W(T^{-1}+\xi)}\left\{\left[2-e^{(W+V_b)\xi}\right.\right. \nonumber \\ &\left.\Bigl. -e^{W\xi}\Bigr](\xi^{-1}-T)V_\mr{th}^{-1}+e^{W\xi}\right\},
\end{align}
where $W=(V_\mr{th}-V_b)/2$. In this article we concentrate on the cases where along the semiclassical trajectory, described by Eq.~\eqref{eq:semiclassical}, $\xi\ll1/V_C,\;1/T_C$. In this case, a straightforward expansion to second order in $\xi$ yields
\begin{align}
 {\cal S}_{\mr{se}}=&-g_T TV_\mr{th}^{-1}e^{-W/T}\left\{\left[W^2+2TW+TV_b-\frac{V_b^2}{2}\right]\xi\right. \nonumber \\ &\left.-\left[\frac{W^3}{3}+TW^2-\frac{TV_b^2}{2}+\frac{V_b^3}{6}\right]\xi^2\right\}.
\end{align}
Near the critical bias voltage, $V_b\approx V_C=(\sqrt{2}-1)V_\mr{th}$, it is convenient to work with dimensionless variables
\begin{equation*}
 t=\sqrt{2}T/V_C,\;v=(V_b-V_C)/V_C,\;x=\xi V_C,
\end{equation*}
so that the action can be expanded to
\begin{equation}
 \mathcal{S}_{\mr{se}}=-g_T t e^{-\frac{1-v/\sqrt{2}}{t}}\left((t-v)x-\frac{x^2}{6}\right)\frac{V_C}{2},
\end{equation}
valid when $t\ll1,\;|v|\ll1,\;|x|\ll1$. From this form we can immediately see that the relevant $x\approx6(t-v)\ll1$, justifying our expansion above.

The cotunneling part of the action can be written as\cite{laakso10}
\begin{align}
\label{eq:cotaction}
 {\cal S}_{\mr{cot}}=&\dot{H}_\mr{cot}\xi+\frac{1}{2}S_{\dot{H},\mr{cot}}\xi^2, \nonumber \\ \approx&\alpha\frac{g_T^2V_C^2\xi}{2}+\beta\frac{g_T^2V_C^3\xi^2}{12},
\end{align}
where $\dot{H}_\mr{cot}$ and $S_{\dot{H},\mr{cot}}$ are the energy current and its fluctuations due to cotunneling, and $\alpha$ and $\beta$ numerical factors of the order of $0.1$. Here we disregard the weak voltage and temperature dependence of these terms. It turns out that the energy current noise due to cotunneling, i.e., the term proportional to $\beta$, can be neglected: The single-electron term is proportional to $g_T t\exp(-(1-v/\sqrt{2})/t)$, whereas the cotunneling term is proportional to $g_T^2$. Near the crossover the exponential term is of the order of $g_T t^{-3}$ (see Eq.~\eqref{eq:t0}) and the single-electron contribution is therefore $g_T^2 t^{-2}\gg g_T^2$.

The total action, including single-electron and cotunneling contributions with the aforementioned approximations, is then
\begin{equation}\label{eq:action}
 \mathcal{S}=-g_T t e^{-\frac{1-v/\sqrt{2}}{t}}\left((t-v)x-\frac{x^2}{6}\right)\frac{V_C}{2}+\alpha\frac{g_T^2xV_C}{2}.
\end{equation}
Using Eq.~\eqref{eq:energy} we substitute
\begin{equation}
 \xi\mapsto-\frac{\partial}{\partial E}=\frac{6\delta_I}{\pi^2tV_C^2}\frac{\partial}{\partial t},
\end{equation}
and obtain for the distribution of temperature,
\begin{equation}
 \mathcal{P}(t,\tau)=\frac{\pi^2V_C^2t}{6\delta_I}\mathcal{P}\left(E=\frac{\pi^2V_C^2t^2}{12\delta_I},\tau\right),
\end{equation}
the Fokker--Planck equation
\begin{align}
 \frac{\pi^2}{3\delta_I g_T}\frac{\partial}{\partial\tau}\mathcal{P}(t,\tau)=-\frac{\partial}{\partial t}\left\{D_1(t)\mathcal{P}(t,\tau)-D_2(t)\frac{\partial}{\partial t}\mathcal{P}(t,\tau)\right\},
\end{align}
where
\begin{align*}
 D_1(t)&=-(t-v)e^{-\frac{1-v/\sqrt{2}}{t}}+\frac{\alpha g_T}{t}-\frac{\delta_I}{\pi^2V_C}\frac{1-v/\sqrt{2}}{t^3}e^{-\frac{1-v/\sqrt{2}}{t}}, \\
 D_2(t)&=\frac{\delta_I}{\pi^2V_C}\frac{1}{t}e^{-\frac{1-v/\sqrt{2}}{t}},
\end{align*}
which, using Eq.~\eqref{eq:stationary}, yield the stationary distribution
\begin{widetext}
\begin{align}\label{eq:stdist1}
 \mathcal{P}_\mr{st}(t)\propto\exp\left\{-\frac{\pi^2V_C}{\delta_I}\left[\frac{1}{3}t^3-\frac{1}{2}t^2v-\alpha g_T t\exp\left(\frac{1-v/\sqrt{2}}{t}\right)+\alpha g_T(1-v/\sqrt{2})\mr{Ei}\left(\frac{1-v/\sqrt{2}}{t}\right)\right]+\frac{1-v/\sqrt{2}}{t}\right\}.
\end{align}
\end{widetext}
Here, $\mr{Ei}(z)$ is the exponential integral function.\cite{weisstein}

To focus on the crossover regime, we introduce a new set of dimensionless parameters by changing the variables to 
\begin{align*}
 t&=\sqrt{2}T_C/V_C+2\theta(T_C/V_C)^2,\\ v&=\sqrt{2}T_C/V_C+2\nu(T_C/V_C)^2,
\end{align*}
with the temperature at the crossover, $T_C$, defined through Eq.~\eqref{eq:t0}, see also Appendix \ref{sec:rescaling}. We also introduce a dimensionless level spacing,
\begin{equation*}
 \mathcal{D}=\frac{1}{4\sqrt{2}\pi^2}\frac{\delta_I}{eV_C}\left(\frac{eV_C}{k_BT_C}\right)^5,
\end{equation*}
characterizing the relative strength of the fluctuations, and a typical relaxation time for these fluctuations in the crossover interval,
\begin{equation*}
 \tau_r=\frac{2\sqrt{2}\pi^2}{3\alpha g_T^2}\frac{\hbar}{\delta_I}\left(\frac{k_BT_C}{eV_C}\right)^3.
\end{equation*}
For a typical $g_T=10^{-3}$, $\tau_r\approx1.7\times10^{-10}/\delta_I\:\mr{eVs}$. When $\nu,\;\theta\ll V_C/T_C$, and $T_C/V_C\ll1$, we can expand in the small parameter, $T_C/V_C$, and simplify the Fokker--Planck equation for the overheated SET considerably:
\begin{align}
 \tau_r\frac{\partial}{\partial\tau}\mathcal{P}(\theta,\tau)=&-\frac{\partial}{\partial \theta}\biggl\{\left[-(\theta-\nu)e^\theta+1-\mathcal{D}e^\theta\right]\mathcal{P}(\theta,\tau)\biggr. \nonumber \\ &\left.-\mathcal{D}e^\theta\frac{\partial}{\partial \theta}\mathcal{P}(\theta,\tau)\right\}.
\end{align}

The stationary distribution becomes
\begin{equation}\label{eq:stdist2}
 \mathcal{P}_\mr{st}(\theta)\propto\exp\left\{-\frac{1}{\mathcal{D}}\left(\frac{1}{2}\theta^2-\nu
     \theta+e^{-\theta}\right)-\theta\right\},
\end{equation}
which is the main result of this paper. The distribution consists of two parts: the first, proportional to $1/\mathcal{D}$, which is also present in the semiclassical (saddle-point) limit, and the second, independent of $\mathcal{D}$, describing the Gaussian fluctuations around the saddle-point trajectories, which become significant as $\delta_I$ (and therefore $\mathcal{D}$) grows. Low-temperature fluctuations are exponentially suppressed since the cooling due to singe-electron tunneling is also exponentially small. High-temperature fluctuations on the other hand do not have such suppressing mechanisms, hence they follow the usual Gaussian form.

\section{Regime of small fluctuations}\label{sec:small}

In the limit of a metallic island, $\mathcal{D}\ll1$, the last terms in the exponents in Eqs.~\eqref{eq:stdist1} and \eqref{eq:stdist2} can be disregarded. In this limit the typical fluctuations are Gaussian and small: the variance is proportional to $\delta_I/V_C$. It should be stressed that our approach still goes beyond the Gaussian approximation -- we can also look at \textit{atypical}, large fluctuations, which are non-Gaussian in character. 

On the single-electron side of the crossover, but yet not far from the crossover point, $1\gg v\gg T_C/V_C$, we can in addition neglect the exponential terms, proportional to $g_T$, in Eq.~\eqref{eq:stdist1}:
\begin{equation}
 \mathcal{P}_\mr{st}(t)\propto\exp\left\{-\frac{\pi^2V_C}{\delta_I}\left[\frac{1}{3}t^3-\frac{1}{2}t^2v\right]\right\}.
\end{equation}
The distribution has a maximum at $t=v$. The cubic term favors low-temperature fluctuations, similar to that obtained in \refcite{heikkila09} for the noninteracting voltage biased island.\footnote{We note that the axis labels in Fig.~3 of \refcite{heikkila09} are wrong.} This is because far above the crossover SE processes serve as an efficient cooling mechanism. Similar to the results in \refcite{heikkila09}, deviations from Gaussian statistics appear for temperature fluctuations of the order of average temperature, but their probability is greatly enhanced since $\ln\mathcal{P}\simeq T^3/\delta_I V_C^2 \ll T/\delta_I$.

In the competition regime, $v<0$, we can approximate the exponential integral with its asymptotic form, $\mr{Ei}(z)\approx e^z(1/z+1/z^2)$, for large $z$. The resulting distribution is
\begin{align}
 \mathcal{P}_\mr{st}(t)\propto\exp&\left\{-\frac{\pi^2V_C}{\delta_I}\left[\frac{1}{3}t^3-\frac{1}{2}t^2v\right.\right. \nonumber \\ &\left.\left.+\alpha g_T \frac{t^2}{1-v/\sqrt{2}}\exp\left(\frac{1-v/\sqrt{2}}{t}\right)\right]\right\}.
\end{align}
In contrast to the previous case, low-temperature fluctuations are exponentially suppressed for reasons explained at the end of Sec.~\ref{sec:theory}.

Around the crossover, $v\lesssim T_C/V_C$, we have from Eq.~\eqref{eq:stdist2}
\begin{equation}
 \mathcal{P}_\mr{st}(\theta)\propto\exp\left\{-\frac{1}{\mathcal{D}}\left(\frac{1}{2}\theta^2-\nu
     \theta+e^{-\theta}\right)\right\}.
\end{equation}
This distribution is ``half-Gaussian'', i.e., Gaussian for high-temperature fluctuations but exponentially suppressed for low-temperature fluctuations. In this case the non-Gaussian character appears already for deviations of temperature of the order of $\delta T\simeq (T_C/V_C)T_C$ and their probability is even more enhanced, $\ln\mathcal{P}\simeq (T_C/V_C)^4(T_C/\delta_I)$.

\begin{figure}
 \includegraphics[width=\columnwidth]{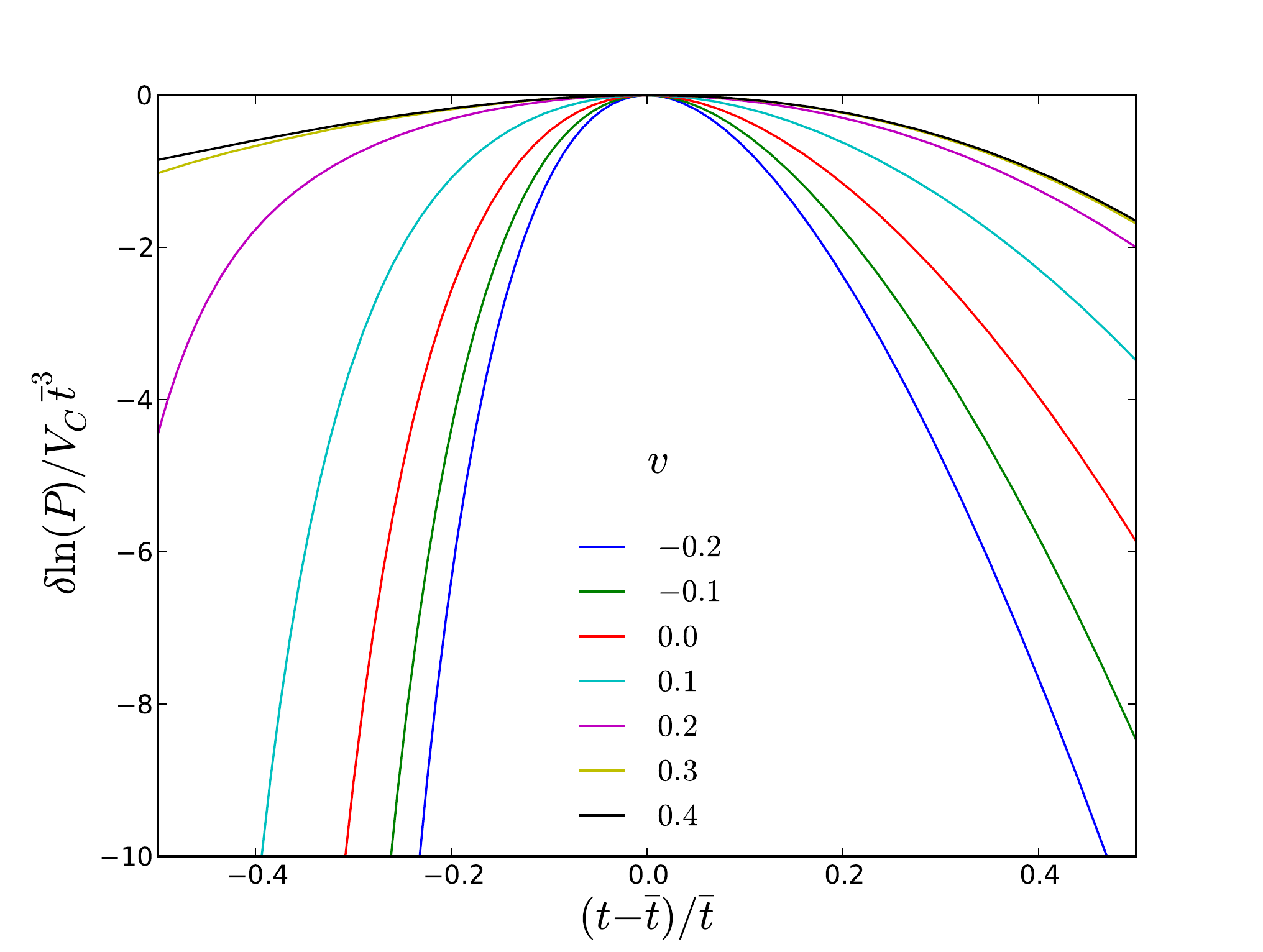}
 \caption{(color online) Logarithm of the temperature fluctuation probability for some values of bias voltage $v=(V_b-V_C)/V_C$. In this plot $\alpha=0.1$ and $g_T=10^{-3}$.}
 \label{fig:logdist_t}
\end{figure}

\begin{figure}
 \includegraphics[width=\columnwidth]{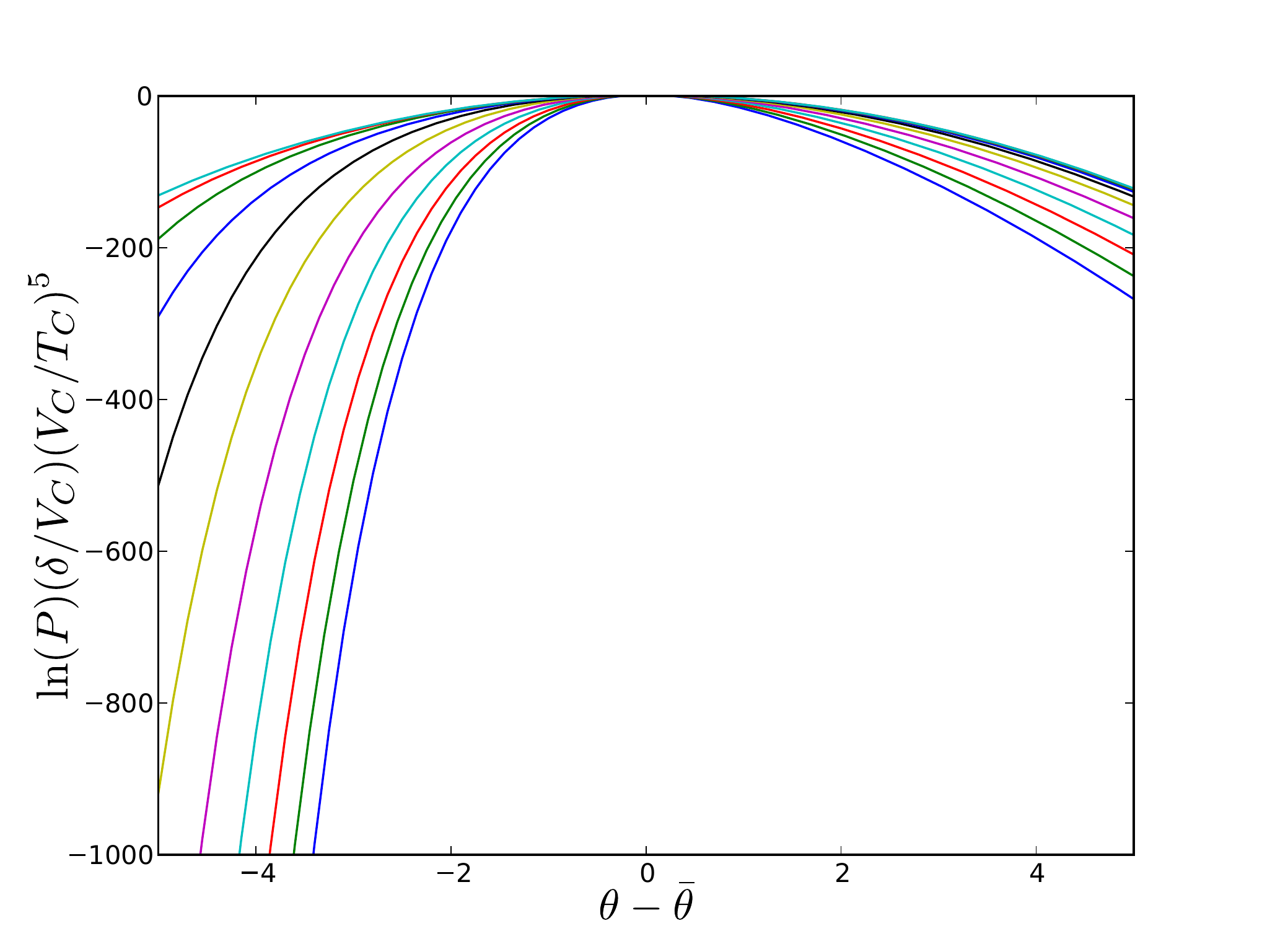}
 \caption{(color online) Logarithm of the temperature fluctuation probability for bias voltages $\nu=-5\ldots5$, bottom to top.}
 \label{fig:logdist_y}
\end{figure}

The logarithm of these probability distributions is plotted in Figs.~\ref{fig:logdist_t} and \ref{fig:logdist_y}, which clearly show the half-Gaussian characteristics for voltages close to and below $V_C$.

\section{Giant fluctuations}\label{sec:large}

In the previous section, we assumed that the fluctuations are small, $\mathcal{D}\ll1$. If $\mathcal{D}\simeq 1$, the typical fluctuations become non-Gaussian. For temperature, the fluctuations are still small, occurring at the scale of $\delta T\simeq T_C(T_C/V_C)$. Since current is
anomalously sensitive to temperature, these fluctuations manifest in much stronger fluctuations of the current. Throughout this section we work with the scaled variables, $\nu$ and $\theta$, and the results presented are valid for $\nu,\;\theta\ll V_C/T_C$.

\subsection{Temperature fluctuations}
\begin{figure}
 \centering
 \includegraphics[width=\columnwidth]{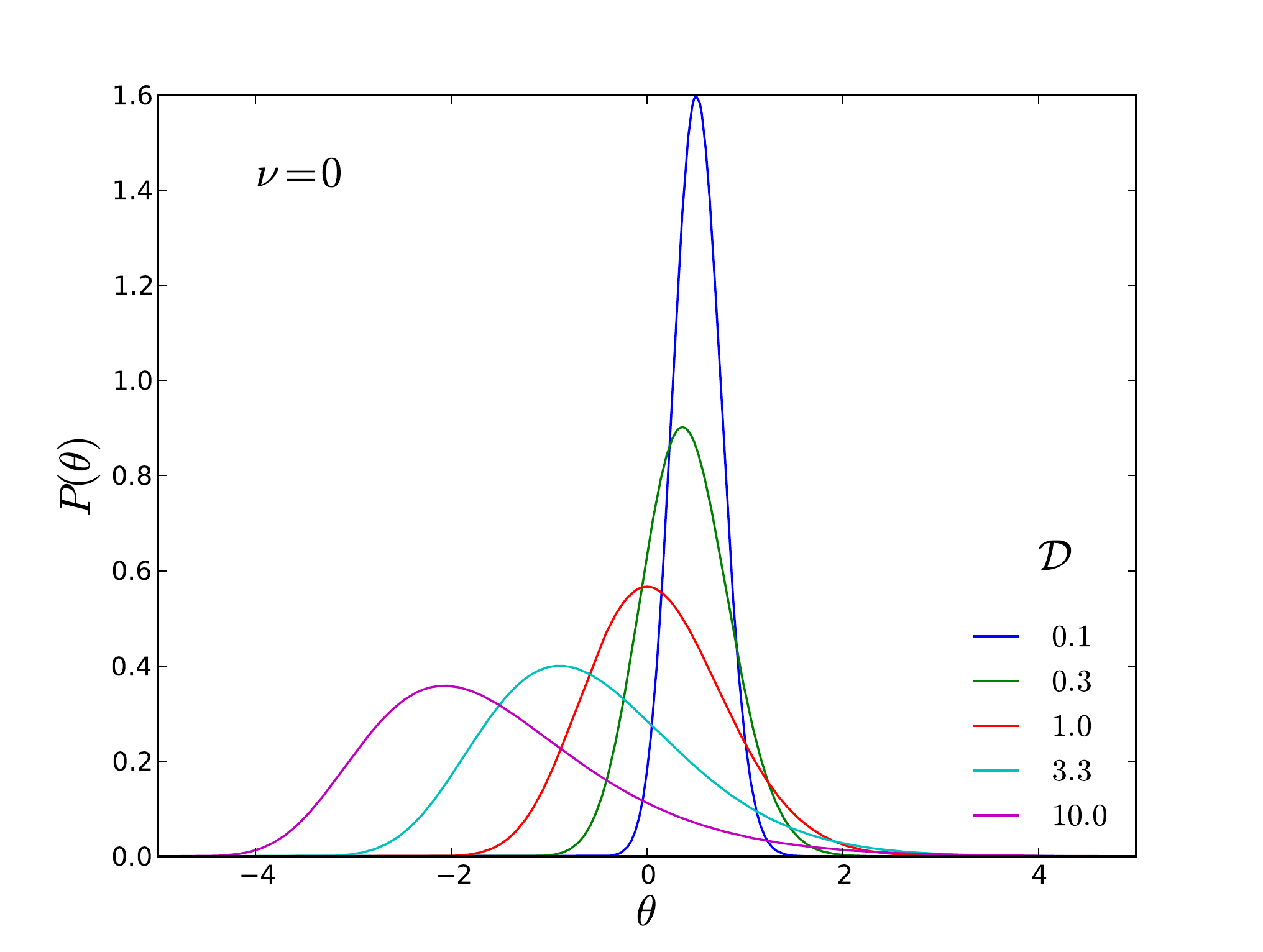}
 \caption{(color online) Probability distribution for the temperature of the island. Note how the maximum of the distribution is shifted and the non-Gaussian tail becomes more prominent as $\mathcal{D}$ is increased.}
 \label{fig:set-scaled}
\end{figure}

\begin{figure}
 \centering
 \includegraphics[width=\columnwidth]{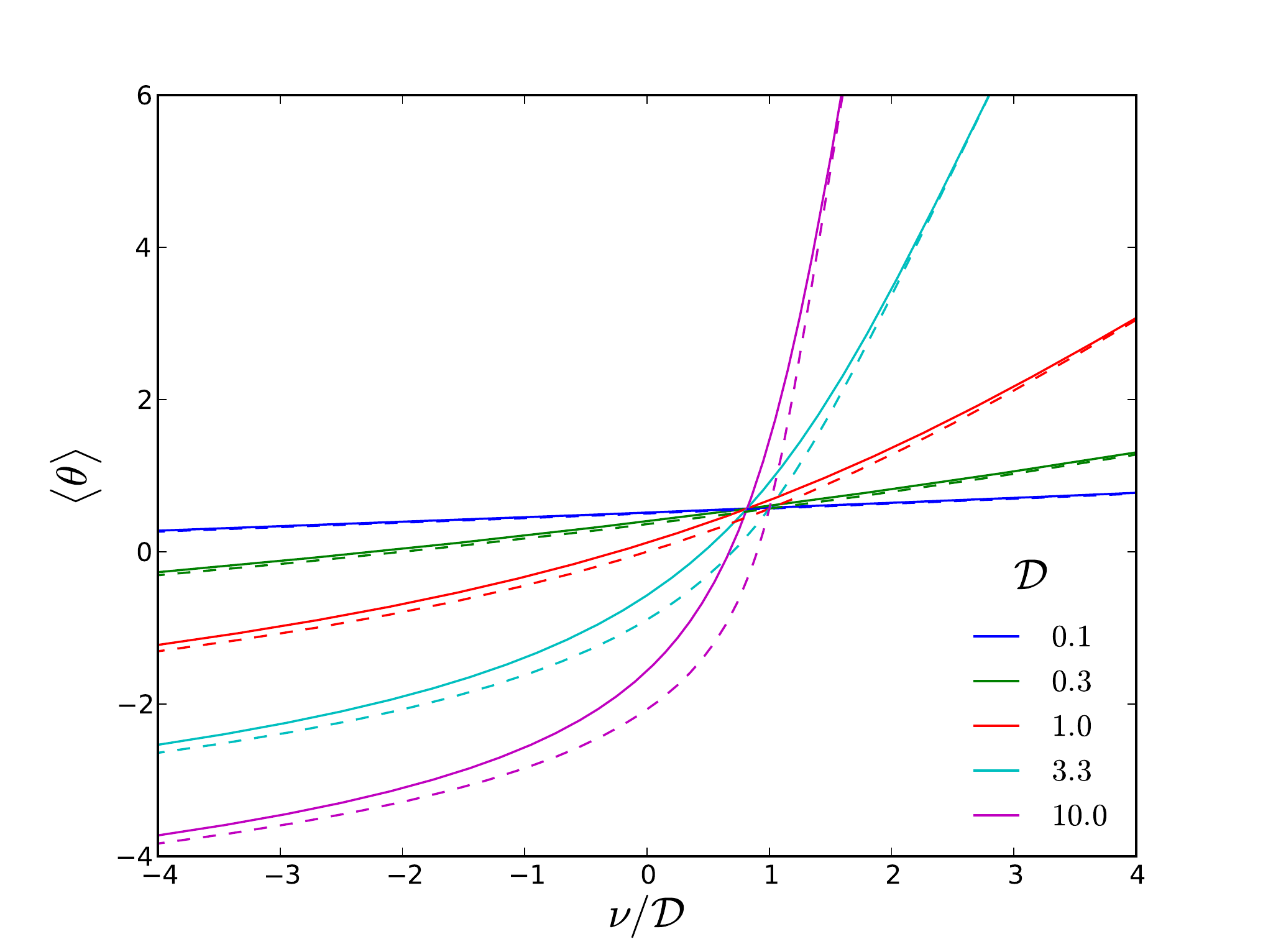}
 \caption{(color online) Expectation value of the temperature as a function of bias voltage. Dashed lines show the mode of $\theta$, i.e., the maximum of the distribution. The two tend to each other for $|\nu| \gtrsim \mathcal{D}$.}
 \label{fig:avgy}
\end{figure}

\begin{figure}
 \centering
 \includegraphics[width=\columnwidth]{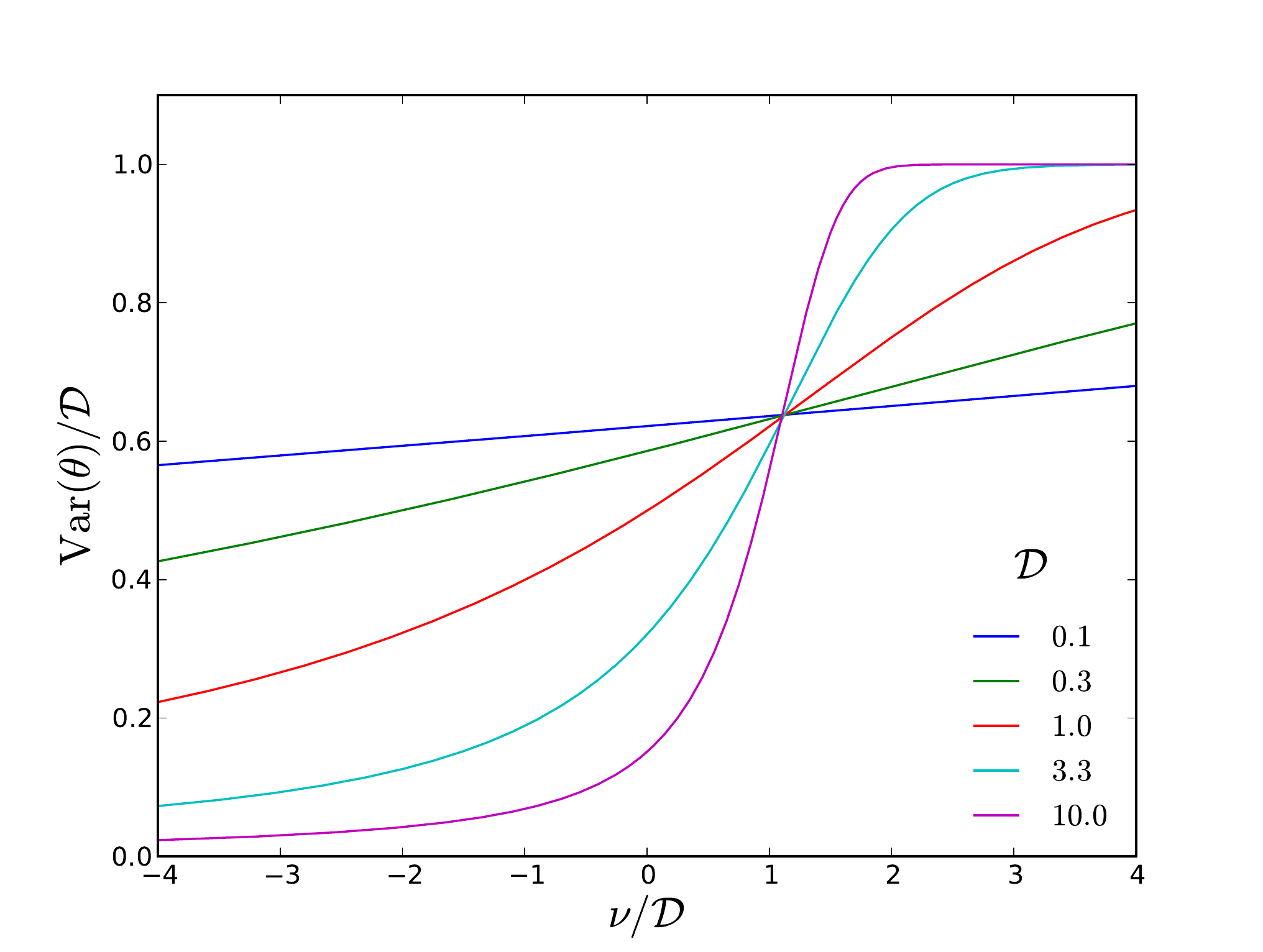}
 \caption{(color online) Variance of the temperature as a function of bias voltage. For $\nu/\mathcal{D}\gg1$ the variance approaches $\mathcal{D}$.}
 \label{fig:vary}
\end{figure}

For a finite but small level spacing $\delta_I$, the distribution of temperature is given by Eq.~\eqref{eq:stdist2} and plotted in Fig.~\ref{fig:set-scaled}. The maximum of the distribution is located at 
\begin{equation}\label{eq:modey}
 \theta_{\max}=\left\{\begin{array}{cc}
           \nu-\mathcal{D},&\; \nu>0\\
	   -\ln\left(-\nu+\mathcal{D}\right),&\; \nu<0
          \end{array}\right.,\;|\nu|\gg\mathcal{D}.
\end{equation}
The maximum is shifted to lower temperatures with increasing $\mathcal{D}$ and it differs from the expectation value of the temperature when $|\nu|/\mathcal{D}$ is small as shown in Fig.~\ref{fig:avgy}. The variance of $\theta$ is plotted in Fig.~\ref{fig:vary}.

Analytical approximations for the expectation value and variance can be obtained in the limit $|\nu|\gg\mathcal{D}$, $\nu>0$ and $\nu<0$. In the first case the weight of the distribution is shifted to large $\theta$, and we can neglect the term $e^{-\theta}$ in Eq.~\eqref{eq:stdist2}. In the latter case the weight is at small $\theta$, and we can neglect the quadratic term in Eq.~\eqref{eq:stdist2}. We obtain
\begin{align}
 \langle\theta\rangle&=\left\{\begin{array}{cc}
                               \nu-\mathcal{D},&\; \nu>0 \\
                               -\ln(-\nu),&\; \nu<0
                              \end{array}\right.,\;|\nu|\gg\mathcal{D}, \\
 \mr{Var}(\theta)&=\left\{\begin{array}{cc}
                           \mathcal{D},&\; \nu>0 \\
                           -\mathcal{D}/\nu,&\; \nu<0
                          \end{array}\right.,\;|\nu|\gg\mathcal{D}.
\end{align}
For large positive or negative $\nu$, $\langle\theta\rangle\approx\theta_{\max}$. Variance is proportional to $\mathcal{D}$ as in the regime of small fluctuations.

\subsection{Induced current fluctuations}

\begin{figure}
 \centering
 \includegraphics[width=\columnwidth]{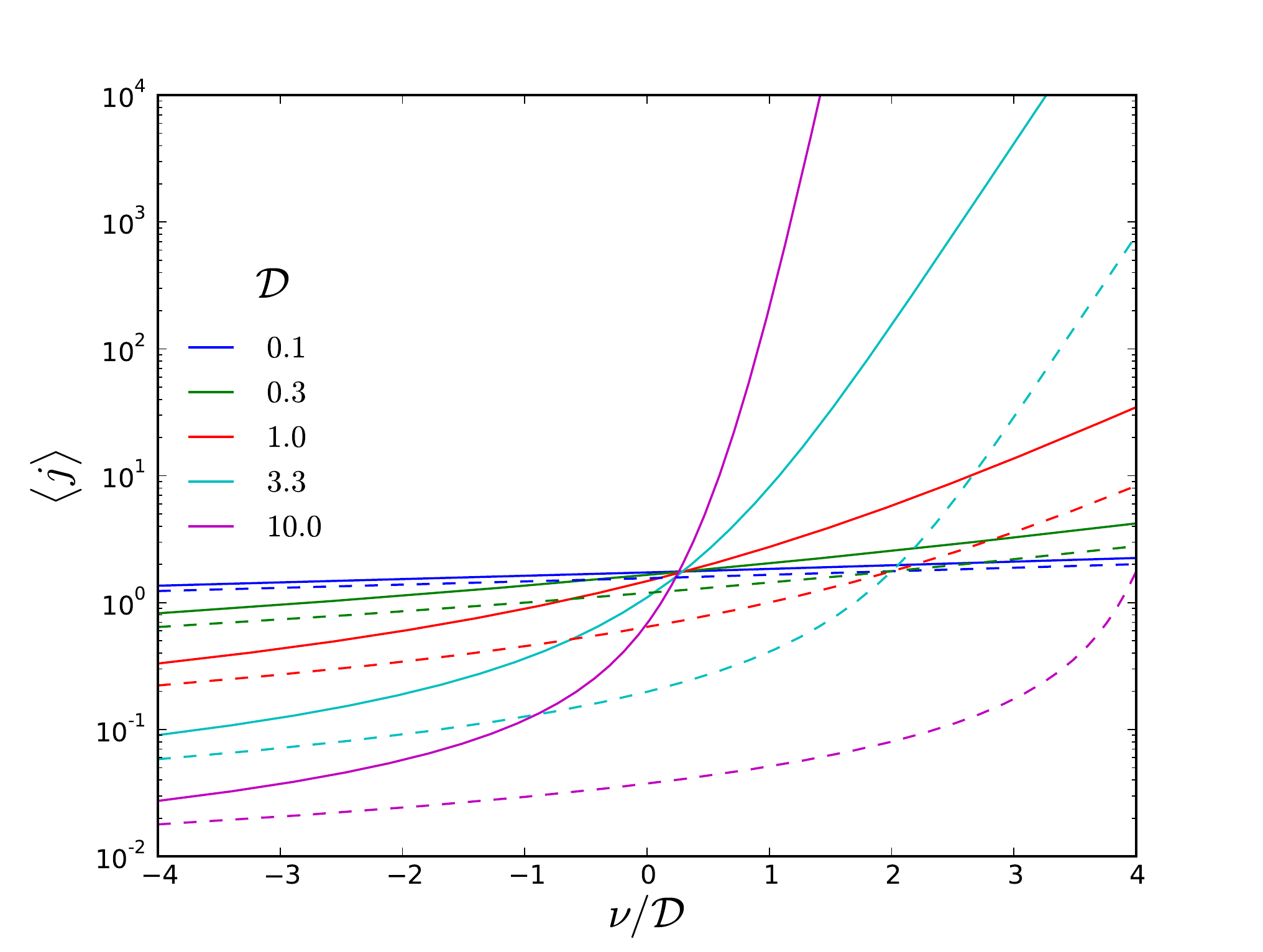}
 \caption{(color online) Expectation value of the current as a function of bias voltage. Dashed lines show the mode of $j$, i.e., the maximum of the distribution. The ratio of these tend to $\exp(3\mathcal{D}/2)$ for $\nu/\mathcal{D}\gg1$, $\mathcal{D}\gg1$.}
 \label{fig:avgj}
\end{figure}

\begin{figure}
 \centering
 \includegraphics[width=\columnwidth]{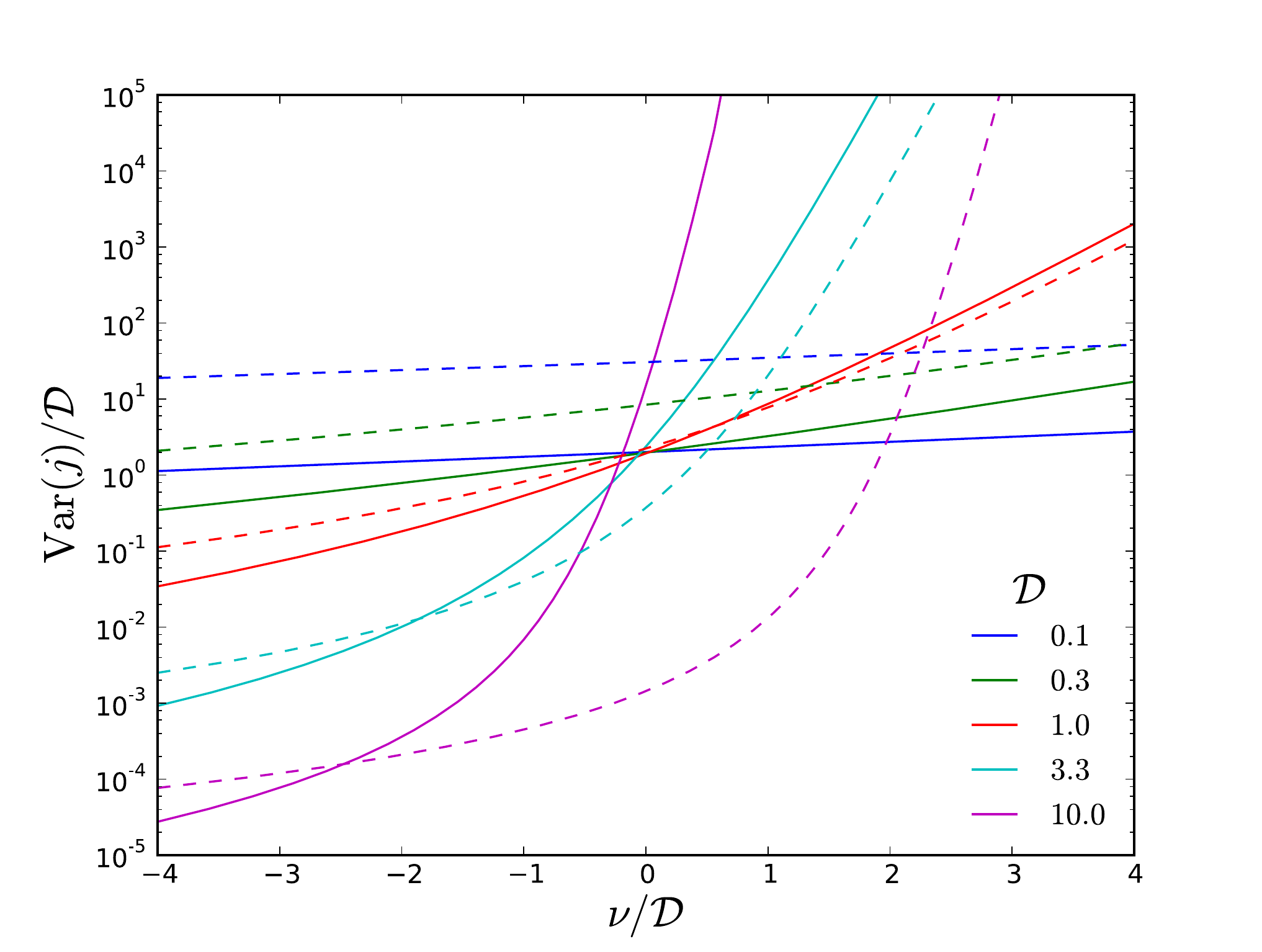}
 \caption{(color online) Variance of the current as a function of bias voltage. Dashed lines show the square of the average current, $\langle j\rangle^2/\mathcal{D}$, for comparison. The ratio of these tend to $\exp(\mathcal{D})$ for $\nu/\mathcal{D}\gg1$, $\mathcal{D}\gg1$.}
 \label{fig:varj}
\end{figure}

In the overheated SET the electric current near the critical bias is given by\cite{laakso10}
\begin{equation}
 I(\theta)=\frac{e^2}{\hbar}\frac{\alpha g_T^2e^2V_C^3}{4k_B^2T_C^2}e^\theta.
\end{equation}
A fluctuating temperature leads directly to a fluctuating electric current. Since the current depends exponentially on the temperature, small fluctuations of temperature lead to large fluctuations of current. The Fokker--Planck equation for temperature is easily converted to a corresponding equation for the dimensionless current, $j\equiv e^\theta$, with the prescription
\begin{equation*}
 \frac{\partial}{\partial\theta}=j\frac{\partial}{\partial j},\;\mathcal{P}(j,t)=\frac{1}{j}\mathcal{P}(\theta=\ln j,t),
\end{equation*}
resulting in
\begin{align}
 \tau_r\frac{\partial}{\partial t}\mathcal{P}(j,t)=&-\frac{\partial}{\partial j}\biggl\{j\left[j\left(\nu-\ln j-2\mathcal{D}\right)+1\right]\mathcal{P}(j,t)\biggr.\nonumber \\ &\left.-\mathcal{D}j^3\frac{\partial}{\partial j}\mathcal{P}(j,t)\right\}.
\end{align}
Note that this equation only includes the dominating contribution to the current fluctuations due to the temperature fluctuations, and the intrinsic fluctuations (thermal and shot noise) are disregarded. The stationary distribution of current is then 
\begin{equation}\label{eq:jdist}
 \mathcal{P}_\mr{st}(j)\propto\exp\left\{-\frac{1}{\mathcal{D}}\left(\frac{1}{2}(\ln j)^2-\nu\ln j+\frac{1}{j}\right)-2\ln j\right\}.
\end{equation}
Similarly to the case of temperature, the maximum of the distribution is shifted to lower values of $j$ when $\mathcal{D}$ is increased. The maximum of the distribution is located at
\begin{equation}\label{eq:modej}
 j_{\max}=\left\{\begin{array}{cc}
           \exp\left(\nu-2\mathcal{D}\right),&\; \nu>0 \\
           \left(-\nu+2\mathcal{D}\right)^{-1},&\; \nu<0
          \end{array}\right.,\;|\nu|\gg2\mathcal{D}.
\end{equation}
Due to the log normal character of the current distribution, the expectation value and the most probable value of the current deviate as $\nu$ is increased. This is shown in Fig.~\ref{fig:avgj}. For a small $\mathcal{D}$, this deviation is proportional to $\mathcal{D}$. The variance of $j$ is plotted as a function of the (reduced) bias voltage in Fig.~\ref{fig:varj}. Note that $\mr{Var}(j)$ is the variance of the instantaneous electric current, which is different from the zero frequency spectral noise power, more often encountered in the literature.

To comprehend the unusual properties of the distribution, let us first turn to the limit $\mathcal{D}\gg 1$, $|\nu|/\mathcal{D}\simeq 1$. In this case, the distribution reduces to a power-law function
\begin{equation}
\mathcal{P}_\mr{st}(j)\simeq\frac{1}{j^{2-\nu/\mathcal{D}}}.
\end{equation}
This form cannot be valid at all values of the current: It is cut off at small currents, $j_s\simeq 1/2\mathcal{D}$, and crosses over to the log-normal distribution at large currents, $j_b\simeq \exp(\mathcal{D})$, $j_s \ll 1 \ll j_b$.
% For $\nu/\mathcal{D}<2$, the power-law function is decreasing, and we can estimate the maximum of the distribution to lie at $j_{\max}\simeq j_s$. Similarly, for $\nu/\mathcal{D}>2$, the power-law function is increasing, and $j_{\max}\simeq j_b$. Estimating the expectation value and variance of the current with a distribution limited between $j_s$ and $j_b$ yields
% \begin{align}
% \langle{j}\rangle&\simeq\left\{\begin{array}{cc}
%                                 j_s,\;&\nu/\mathcal{D}<0, \\ 
%                                 j_b,\;&\nu/\mathcal{D}>0,
%                                \end{array}\right. \\
% \sqrt{\mr{Var}(j)}&\simeq\left\{\begin{array}{cc}
%                                 j_s,\;&\nu/\mathcal{D}<-1, \\ 
%                                 j_b,\;&\nu/\mathcal{D}>-1.
%                                \end{array}\right.
% \end{align}
% We notice that $\langle j \rangle$ parametrically exceeds $j_\mr{max}$ for $0<\nu$, and $\sqrt{\mr{Var}(j)}$ parametrically exceeds $\langle j \rangle$ for $-\mathcal{D}<\nu$. This signals a highly unusual distribution not satisfying, for instance, the central limit theorem conditions.
% 
% We stress that the exponent of the power-law distribution can be readily tuned with bias voltage, $\d V_b=2V_C(T_C/V_C)\d \nu$. We are not aware of any other physical systems where a quantity exhibits a power-law distribution with a tunable exponent.
Analytical approximations for the expectation value and variance can be obtained in the limit $|\nu|\gg\mathcal{D}$, $\nu>0$ and $\nu<0$. Similarly to the case of the temperature distribution, in the first case the weight of the distribution is shifted to large $j$, and we can neglect the term $1/j$ in Eq.~\eqref{eq:jdist}. In the latter case we can neglect the $(\ln j)^2$ term. We obtain
\begin{align}\label{eq:avgj}
 \langle j\rangle&=\left\{\begin{array}{cc}
                           \exp(\nu-\mathcal{D}/2),&\; \nu>0\\
                           -1/\nu,&\; \nu<0\\
                          \end{array}\right.,\;|\nu|\gg\mathcal{D}, \\
 \mr{Var}(j)&=\left\{\begin{array}{cc}\label{eq:varj}
                      \exp(2\nu-\mathcal{D})(\exp(\mathcal{D})-1),&\; \nu>0 \\
                      \mathcal{D}\left[\nu^2(-\nu-\mathcal{D})\right]^{-1},&\; \nu<0
                     \end{array}\right.,\;|\nu|\gg\mathcal{D}.
\end{align}
We notice that for $\nu>0$, $\langle j \rangle$ parametrically exceeds $j_\mr{max}$, $\langle j \rangle/j_\mr{max}=\exp(3\mathcal{D}/2)$, and $\sqrt{\mr{Var}(j)}$ parametrically exceeds $\langle j \rangle$, $\sqrt{\mr{Var}(j)}/\langle j \rangle=\exp(\mathcal{D}/2)$. This signals a highly unusual distribution not satisfying, for instance, the central limit theorem conditions.

We stress that the exponent of the power-law distribution can be readily tuned with bias voltage, $\d V_b=2V_C(T_C/V_C)\d \nu$. We are not aware of any other physical systems where a quantity exhibits a power-law distribution with a tunable exponent.

\begin{figure}
 \centering
 \includegraphics[width=\columnwidth]{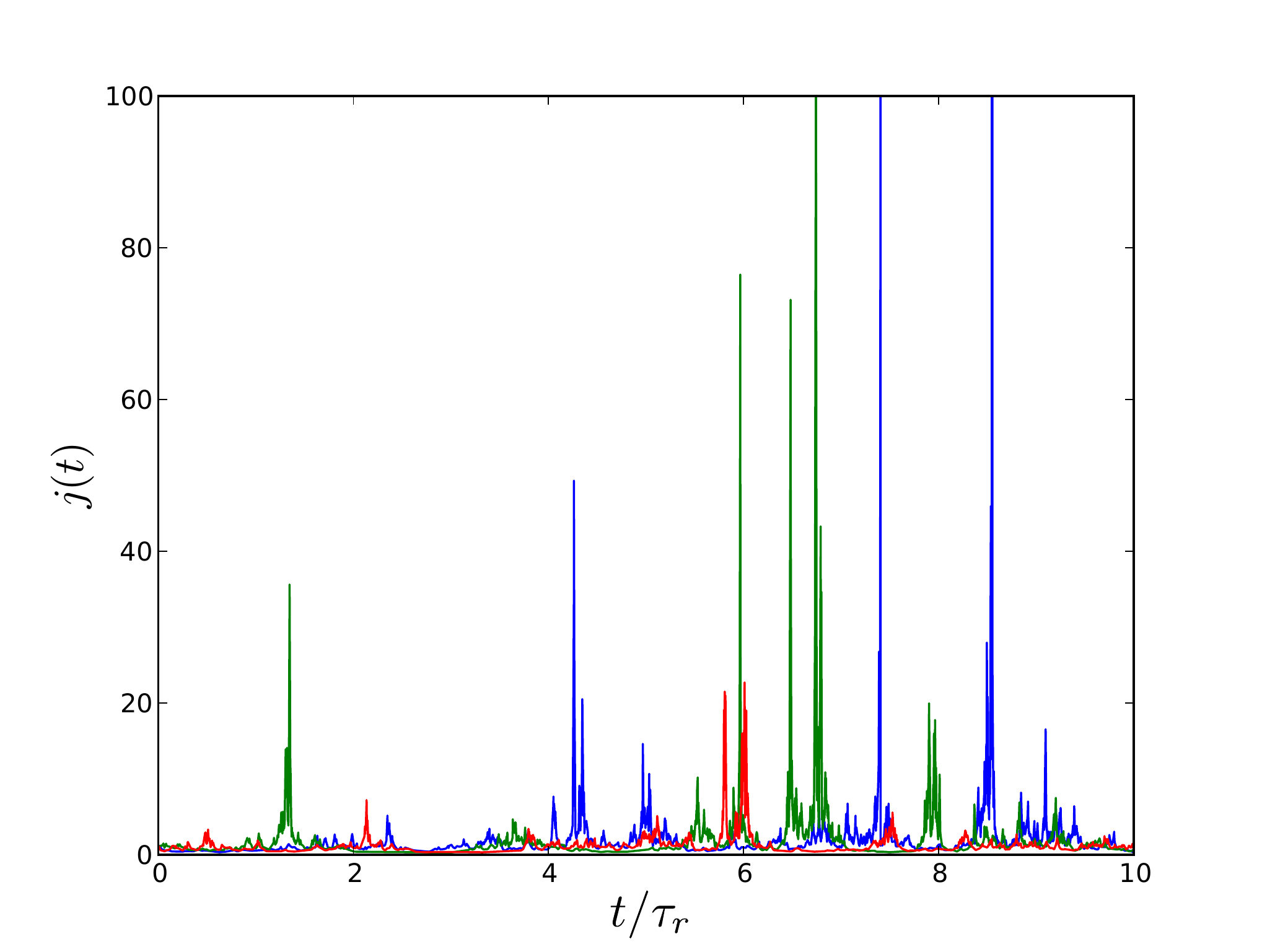}
 \caption{(color online) Three simulated realizations of current in an overheated SET at $\nu=0$, $\mathcal{D}=3.3$. The average current (for $\mathcal{D}\to0$) is $j\approx1.7$.}
 \label{fig:current_time}
\end{figure}

Using Eq.~\eqref{eq:langevin} allows us to write down a Langevin equation for the time dependence of the current
\begin{equation}
 \tau_r\frac{\partial}{\partial t}j(t)+\gamma(j(t))j(t)=\eta(t),
\end{equation}
where $\gamma(j)=-j\left(\nu-\ln j+\mathcal{D}\right)-1$ and $\eta(t)$ satisfies $\langle\eta(t)\rangle=0$ and $\langle\eta(t)\eta(t')\rangle=2\mathcal{D}j(t)^3\delta(t-t')$. The current exhibits huge peaks at a low frequency that are orders of magnitude larger than the average current, as can be seen from Fig.~\ref{fig:current_time}. The time scale for these fluctuations, $\tau_r$, is of the order of milliseconds for $g_T=10^{-3}$ and $\delta_I=0.01\:\mr{K}$.

\begin{figure}
 \centering
 \includegraphics[width=\columnwidth]{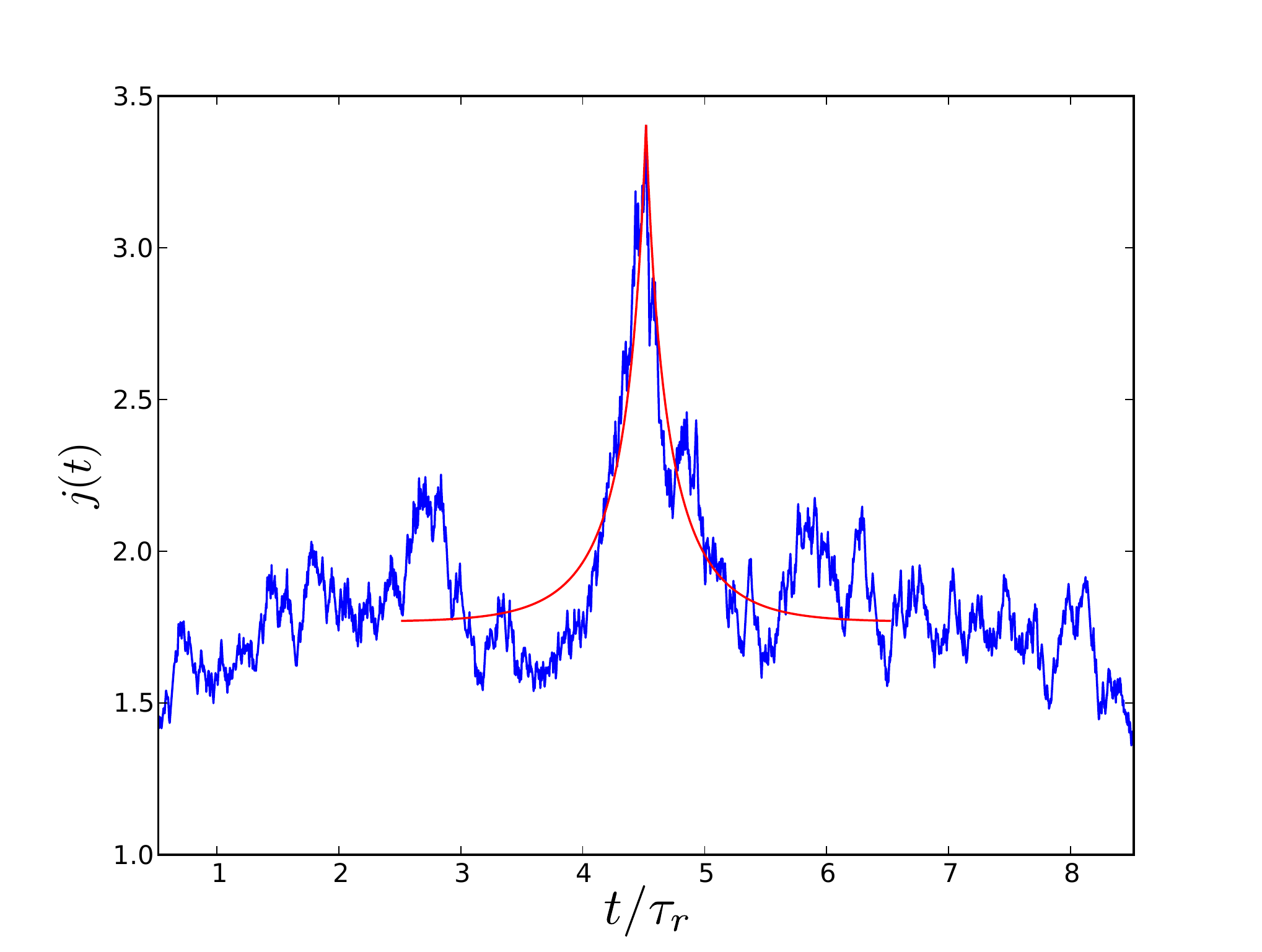}
 \caption{(color online) Comparison of a saddle point fluctuation to the Langevin result at $\nu=0$, $\mathcal{D}=0.02$.}
 \label{fig:comparison}
\end{figure}

Another way to access the time dependence of the current fluctuations is to study the saddle point equations for small fluctuations \eqref{eq:semiclassical}, i.e., vanishing $\delta_I$,
\begin{align}
 \tau_r\dot x&=-\frac{\partial}{\partial \theta}S, \nonumber \\
 \tau_r\dot \theta&=\frac{\partial}{\partial x}S,
\end{align}
where the action,
\begin{equation}
 S=-\left[e^\theta(\theta-\nu)-1\right]x+e^\theta\frac{x^2V_C^2}{12T_C^2},
\end{equation}
has been written in terms of the dimensionless, scaled, variables, $\nu$, $\theta$, and $x$. As explained in \refcite{heikkila09}, the saddle-point solutions follow the trajectory given by $S(\theta(t),x_S(\theta(t)))=0$ with $x_S(\theta) \neq 0$ until a ``measurement'' at time $\tau$ forces them to a trajectory with $x_S \equiv 0$. As shown in Fig.~\ref{fig:comparison}, the shape of a fluctuation obtained from the saddle-point equations agrees with the Langevin result. This is expected, since for small $\mathcal{D}$ the fluctuations around the saddle point trajectory are negligibly small.

\section{Conclusions}\label{sec:conclusion}

We have studied the temperature fluctuations and the associated current fluctuations in an overheated SET. Focusing on the crossover region between the competition of single-electron and cotunneling, and pure sequential tunneling, we have found several interesting features, not seen in other commonly considered nanostructures. In overheated SETs with small islands the expectation value and the most probable value of the temperature differ from each other due to a tail in the probability distribution of temperature, extending to significantly high temperatures. Due to exponential temperature sensitivity of the electric current, huge peaks, occurring at frequencies of the order of a few kilohertz, should be visible in an instantaneous measurement of the electric current.

Experimentally, the challenge is to manufacture a transistor island where, despite high resistance tunnel contacts, the electron--phonon heat current is negligible compared to the heat current to the leads. In earlier experiments on the measurement of noise in SETs (see for example \refcite{kafanov09}), this has not been the case. According to our earlier evaluation\cite{laakso10} a transistor island with a volume of $\mathcal{V}=10^{-4}\:\mu\mr{m}^3$ connected with $g_T=10^{-3}$ tunnel contacts should allow one to detect these large current surges experimentally.

\acknowledgments
This work was supported by the Finnish Academy of Science and Letters, the Academy of Finland, and the European Research Council (Grant No. 240362-Heattronics). 

\appendix
\section{Details on the rescaling}\label{sec:rescaling}
The total energy current in the SET can be found from the action, Eq.~\eqref{eq:action}, via
\begin{equation}
 \dot{H}=\left.\frac{\partial \mathcal{S}}{\partial x}\right|_{x=0}=-g_T t e^{-\frac{1-v/\sqrt{2}}{t}}(t-v)\frac{V_C}{2}+\alpha\frac{g_T^2V_C}{2}.
\end{equation}
The heat balance follows from the requirement for the total energy current to vanish, leading to
\begin{align}
 &-t e^{-\frac{1-v/\sqrt{2}}{t}}(t-v)+\alpha g_T=0, \nonumber \\
 \Leftrightarrow\quad &v=-\alpha g_T t^{-1}e^{\frac{1-v/\sqrt{2}}{t}}+t.
\end{align}
Next we make a change of variables to $t=t_C+\theta t_C^2$ and $v=t_C+\nu t_C^2$, where $t_C$ is for now an arbitrary number. If $\theta,\;\nu\ll1/t_C$, we can expand in these small parameters to obtain
\begin{equation}
 \nu=-\alpha g_T t_C^{-3}e^{\frac{1}{t_C}-\frac{1}{\sqrt{2}}-\theta}+\theta.
\end{equation}
Selecting
\begin{equation}
 \alpha g_T t_C^{-3}\exp\left(\frac{1}{t_C}-\frac{1}{\sqrt{2}}\right)=1 \Rightarrow t_C\approx\frac{1}{\ln\left(1/g_T\right)}\ll1,
\end{equation}
simplifies the heat balance to
\begin{equation}
 \nu=-e^{-\theta}+\theta,
\end{equation}
and also implies $t_C\ll1$, thus justifying the expansion above for a large range of $\nu$ and $\theta$. We define the crossover point from the competition between single-electron and cotunneling to pure single-electron tunneling to be at $\theta=0$, i.e., $t=t_C=\sqrt{2}T_C/V_C$, defining the temperature at the crossover. This happens when $\nu=-1$, i.e., at a voltage of $v=t_C-t_C^2$. This rescaling simplifies also the Fokker--Planck equation, as seen in the main text, since it allows us to approximate
\begin{equation}
 e^{-\frac{1-v/\sqrt{2}}{t}}\approx \frac{\alpha g_T V_C^3}{2\sqrt{2}T_C^3}e^\theta,
\end{equation}
in many of the formulas.

\section{Relation between the Fokker--Planck equation and the functional integral}\label{sec:fp}
\begin{widetext}
 Let us write the FP equation in the form
 \begin{align}
  \frac{\partial}{\partial t}f(q,t)&=-\frac{\partial}{\partial q}\left[D_1(q,t)f(q,t)\right]+\frac{\partial^2}{\partial q^2}\left[D_2(q,t)f(q,t)\right], \nonumber \\
  \frac{\partial}{\partial t}f(q',t)&=\int\d qf(q,t)\left[-D_1(q,t)\frac{\partial}{\partial q}\delta(q'-q)+D_2(q,t)\frac{\partial^2}{\partial q^2}\delta(q'-q)\right],
 \end{align}
 and integrate over time from $t$ to $t+\Delta t$:
 \begin{equation}
  f(q',t+\Delta t)=\int\d qf(q,t)\left[\delta(q'-q)-\Delta tD_1(q,t)\frac{\partial}{\partial q}\delta(q'-q)+\Delta tD_2(q,t)\frac{\partial^2}{\partial q^2}\delta(q'-q)+\mathcal{O}(\Delta t^2)\right].
 \end{equation}
 Now we use the identity $\delta(q'-q)=\int\d p/2\pi \exp(\ii p(q'-q))$ to write
 \begin{equation}
  f(q',t+\Delta t)=\int\d q\frac{\d p}{2\pi}f(q,t)e^{\ii p(q'-q)}\left[1+\ii p\Delta tD_1(q,t)-p^2\Delta tD_2(q,t)+\mathcal{O}(\Delta t^2)\right].
 \end{equation}
 Iterating $N$ times so that $(t_N-t_0)/\Delta t=N$ gives
 \begin{equation}\label{eq:unique}
  f(q_N,t_N)=\prod_{i=0}^{N-1}\int\d q_i\frac{\d p_i}{2\pi}f(q_0,t_0)\exp\left\{\sum_{i=0}^{N-1}\Delta t\left[\ii p_i\frac{q_{i+1}-q_i}{\Delta t}+\ii p_iD_1(q_i,t_i)-p_i^2D_2(q_i,t_i)\right]\right\},
 \end{equation}
 which admits a functional integral representation
 \begin{equation}
  f(q,t)=\int\mathcal{D}q\mathcal{D}pf(q_0,t_0)\exp\left\{\int\d t\left[\ii p\dot{q}+\ii pD_1(q,t)-p^2D_2(q,t)\right]\right\}.
 \end{equation}
 This form of the functional integral is not unique (unlike Eq.~\eqref{eq:unique}), but depends on the discretization procedure.\cite{wissel79} With the substitution $\ii p\mapsto-\xi$, $q\mapsto E$, we get a functional integral in the form of Eq.~\eqref{eq:keldyshZ}. This proves that the FP equation \eqref{eq:fp} corresponds to the partition function of Eq.~\eqref{eq:keldyshZ}.
\end{widetext}

\bibliography{fcs}

\end{document}